\documentclass[aps,twocolumn,groupedaddress,amsmath,amssymb]{revtex4}
\usepackage{amsmath, dsfont}
\usepackage{amssymb}
\usepackage[dvips]{graphics}
\usepackage{epsfig}
\usepackage{calc}
\usepackage{amsfonts}
\usepackage{graphicx}
\usepackage[ansinew]{inputenc}

\begin{document}
\setcounter{page}{1}
\title{New black holes of vacuum Einstein equations with hyperscaling violation and Nil geometry horizons}

\author{Mokhtar Hassa\"ine}
\email{hassaine-at-inst-mat.utalca.cl} \affiliation{Instituto de
Matem\'atica y F\'isica, Universidad de Talca, Casilla 747, Talca,
Chile}

\begin{abstract}
In this paper, we present a new solution of the vacuum Einstein equations in five dimensions which is a static
black hole with hyperscaling violation and with a three-dimensional horizon modeled by one the eight Thurston geometries, namely the Nil geometry. This
homogeneous geometry is non-trivial in the sense that it is neither of constant curvature  nor a product of constant curvature manifolds. Using the Hamiltonian formalism, we identify the mass and entropy of the black hole solution. Curiously enough, in spite of the fact that the entropy turns to be negative, the mass is positive and the first law of thermodynamics holds. We also discuss the extension in higher dimension.
\end{abstract}

\maketitle

%%%%%%%%%%%%%%%%%%%%%%%
\section{Introduction}
%%%%%%%%%%%%%%%%%%%%%%%
An important result on black holes physics is concerned with the Hawking's theorem \cite{Hawking:1971vc} which asserts that the event horizon of an asymptotically flat
stationary four-dimensional black hole obeying the dominant energy condition is a $2-$sphere. The clues of the proof lie in the Gauss-Bonnet
theorem (valid only in two dimensions) and on the energy condition. In view of these restrictions, there are many ways to escape the Hawking's theorem as by
considering higher-dimensional black holes\footnote{There
exists a generalization of the Hawking's theorem in higher dimensions which states that the event horizon admits metrics of positive scalar curvature (positive Yamabe type), \cite{Galloway:2005mf}.} or matter sources that violate the dominant energy condition. In the first case, we can mention the interesting example of
the five-dimensional black ring of Emparan and Reall \cite{Emparan:2001wk}-\cite{Emparan:2001wn} with horizon topology $\mathds{S}^2\times \mathds{S}^1$, while a simple
way to circumvent the dominant energy condition is to add a negative cosmological constant yielding to the Schwarzschild-AdS black hole in dimensions $D\geq 4$. In five dimensions, the event horizon can be
{\it a priori} a compact orientable $3-$dimensional Riemannian manifold. Nevertheless, due to the Thurston geometrization conjecture proved by Perelman, the event horizon can be endowed with a metric
which is locally isometric to one of the eight Thurston geometries \cite{Thurston}. The simplest ones are given by the Euclidean space $\mathds{E}^3$,
the three-sphere $\mathds{S}^3$, the hyperbolic space $\mathds{H}^3$, the products $\mathds{S}^1\times \mathds{H}^2$ and $\mathds{S}^1\times \mathds{S}^2$ with their
standard corresponding metric. In addition, there are three non-trivial homogeneous geometries which
are neither constant curvature nor a product of constant curvature manifolds
called the Nil geometry, the Solv geometry and the geometry of the universal cover of $\mbox{SL}_2(\mathds{R})$ with the following representative metrics (see \cite{Scott} for a nice review)
\begin{subequations}
\label{Thurstongeo}
\begin{eqnarray}
\label{Solvg}
&&\mbox{Solv}:\qquad d\tilde{s}^2=e^{2x_3}dx_1^2+e^{-2x_3}dx_2^2+dx_3^2,\\
\label{Nilg}
&&\mbox{Nil}:\qquad d\tilde{s}^2=dx_1^2+dx_2^2+(dx_3-x_1dx_2)^2,\\
\label{Sl2}
&&\mbox{SL}_2(\mathds{R}):\,\, d\tilde{s}^2= \frac{1}{x_1^2}(dx_1^2+dx_2^2)+\Big(dx_3+\frac{dx_2}{x_1}\Big)^2.
\end{eqnarray}
\end{subequations}
Schematically, these spacetimes can be written as
\begin{eqnarray}
d\tilde{s}^2=\sum_{I=1}^{3}\left({\omega}^I\right)^2,
\end{eqnarray}
where the ${\omega}^I$ are the corresponding left-invariant one-forms. For latter convenience, the capital index $I$ will run as $I=1,2,3$ while we will use the index $i$ for $i=1,2$.

In Ref. \cite{Cadeau:2000tj}, Cadeau and Woolgar have found two interesting families of
static black hole solutions of  Einstein equations with a negative cosmological constant $G_{\mu\nu}+\Lambda g_{\mu\nu}=0$ whose horizon topologies are modeled
by the Solv (\ref{Solvg}) and the Nil (\ref{Nilg}) $3-$geometries. These solutions are generically represented as follows
\begin{eqnarray}
ds^2=-r^{2z}F(r)dt^2+\frac{dr^2}{r^2F(r)}+\sum_{I=1}^{3}a_I\,r^{2q_I}\left({\omega}^I\right)^2,
\label{CW}
\end{eqnarray}
where the $r^{q_I}{\omega}^I$'s are left-invariant one-forms and such that the zero mass metric (which corresponds to $F(r)=1$) is homogeneous. Here, the
$a_I$'s are constants that permit to introduce an eventual additional scale. For the Solv geometry black hole, the set of parameters is given by
\begin{eqnarray}
\label{SolvCW}
&&\mbox{Solv}:\qquad F(r)=1-M/r^3,\\
&&\left\{\Lambda=-9/2, z=1, q_i=1, q_3=0, a_i=1, a_3=2/3\right\},\nonumber
\end{eqnarray}
while for the Nil geometry black hole, we have
\begin{eqnarray}
\label{NilCW}
&&\mbox{Nil}:\qquad F(r)=1-M/{r^{11/2}},\\
&&\left\{\Lambda=-99/8, z=3/2, q_i=1, q_3=2, a_i=1, a_3=11/2\right\}.\nonumber
\end{eqnarray}

The main clue of these constructions lies in the fact that metrics of the form (\ref{CW}) with $F(r)=1$ and with Solv (\ref{Solvg}) or Nil (\ref{Nilg}) geometry base manifolds can describe Einstein spaces $R_{\mu\nu}=-\alpha^2 g_{\mu\nu}$ for a suitable election of the parameters. And then, these solutions can easily be promoted to black hole configurations of the form (\ref{CW}) with a metric function parametrized as
\begin{eqnarray}
F(r)=1-\frac{M}{r^{\sum_I q_I+z}}.
\label{genCW}
\end{eqnarray}
For the higher-dimensional version of the Nil and Solv black holes, this parametrization of the metric function still holds \cite{Hervik:2003vx}.

It is interesting to note that the asymptotic isometries  (or the zero mass isometries) of the Solv and the Nil solutions
contain a dilatation generator whose action on the coordinates is casted as follows
\begin{eqnarray}
t\to \lambda^{z} t,\quad r\to\lambda^{-1}r,\quad x_i\to\lambda x_i,\quad x_3\to \lambda^{q_3} x_3.
\label{Lifsh2}
\end{eqnarray}
Note that in the case of the Solv solution $(z=1, q_3=0)$, there is only one anisotropic direction $x_3$ while for the Nil solution $(z=3/2, q_3=2)$, the anisotropy is reflected
in two directions: one along the time as in the standard Lifshitz case (see below (\ref{Lifsca})) and the other along the coordinate $x_3$. Just to make the discussion self-contained, we recall that the Lifshitz dual metric in arbitrary dimension $D$ is given by \cite{Kachru:2008yh}
\begin{eqnarray}
ds_{{\cal L}}^2=-{r^{2z}}{dt^2}+\frac{dr^2}{r^2}+r^2\sum_{I=1}^{D-2} dx_I^2,
\label{Lifshitzmetric}
\end{eqnarray}
and it enjoys the scaling symmetry with one anisotropic direction
\begin{eqnarray}
t\to \lambda^z t,\quad r\to\lambda^{-1}r,\quad x_I\to\lambda x_I.
\label{Lifsca}
\end{eqnarray}
Hence, in some sense, the class of spacetime metrics (\ref{CW}) with $F=1$ and $z\not=1$ can be refereed as Lifshitz metrics
with two (or more) anisotropic directions. Note that Lifshitz spacetimes
with two (or more) anisotropic directions have been shown to support Abelian massive vector fields in \cite{Taylor:2008tg}.

This analogy between the Nil solution with the Lifshitz spacetimes must be explored further. Recently, there has been some interest in extending the Lifshitz metrics (\ref{Lifshitzmetric}) by introducing an additional parameter, the hyperscaling violation exponent $\theta$, such that the scaling transformations (\ref{Lifsca}) do not act as an isometry but rather
as a conformal transformation. These metrics refereed as  hyperscaling violation metrics are described by the following line element \cite{Charmousis:2010zz}
\begin{eqnarray}
ds_{{\cal H}}^2=\frac{1}{r^{\frac{2\theta}{D-2}}}\left[-{r^{2z}}{dt^2}+\frac{dr^2}{r^2}+r^2\sum_{i=1}^{D-2} dx_i^2\right],
\label{HSVmetric}
\end{eqnarray}
and transform as $ds_{{\cal H}}^2\to \lambda^{\frac{2\theta}{D-2}}ds_{{\cal H}}^2$ under the scaling transformation (\ref{Lifsca}).
Note that this metric is conformally related to the Lifshitz metric (\ref{Lifshitzmetric}) which is recovered in the limiting case $\theta=0$. As mentioned in Refs. \cite{Dong:2012se} and \cite{Alishahiha:2012cm}, these metrics may be of interest in holographic contexts related to condensed matter physics.
Indeed, it has been shown that the hyperscaling violation metric with hyperscaling violation exponent $\theta=D-3$ could be useful to describe a dual theory with an ${\cal O}(N^2)$ Fermi surface where $N$ denotes the number of degrees of freedom. This is strongly suggested by the fact that the holography entanglement entropy (also called von-Neumann entropy, see \cite{Ryu:2006bv}) presents a logarithmic violation of area law, \cite{Huijse:2011ef}. 

As in the Lifshitz case, there
is a physical interest in looking for black holes whose asymptotic behaviors coincide with the hyperscaling violation metric, see e. g. \cite{HyperBH}. These solutions are usually called black holes with hyperscaling violation. 

In the standard AdS/CFT correspondence, it is well-known that the AdS metric which corresponds to (\ref{Lifshitzmetric}) with $z=1$ solves the Einstein equations with a negative cosmological constant. On the other hand,
in order to support Lifshitz spacetimes (\ref{Lifshitzmetric}), the Einstein equations are not sufficient and require the introduction of some matter source like a Proca field or to consider higher-order gravity theories, see e. g. \cite{LifshitzBHs}. Nevertheless, in the case of the hyperscaling violation metric (\ref{HSVmetric}), a simple computation shows that this metric solves the Einstein equations without cosmological constant provided that the dynamical exponent $z$ and the hyperscaling violation exponent $\theta$ are fixed in term of the  dimension $D$ as
\begin{eqnarray}
z=\frac{2(D-2)}{D-3},\qquad \theta=\frac{(D-2)(D-1)}{D-3}.
\label{solGR}
\end{eqnarray}
However, in this case, looking for a non-zero mass black hole solution within the following simple ansatz with a unique metric function $F$
\begin{eqnarray}
ds_{{\cal H}}^2=\frac{1}{r^{\frac{2\theta}{D-2}}}\left[-{r^{2z}}F(r){dt^2}+\frac{dr^2}{r^2 F(r)}+r^2\sum_{i=1}^{D-2} dx_i^2\right],
\label{HSVmetricbh}
\end{eqnarray}
the field equations imply that $F(r)=1$.

Now, from the experience acquired from the work of Cadeau and Woolgar \cite{Cadeau:2000tj} and the analogy observed between their solutions and the Lifshitz metrics, we would like to explore wether hyperscaling violation metric
(\ref{HSVmetric}) with a three-dimensional manifold modeled by one of the non-trivial Thurston geometries (\ref{Thurstongeo}) instead of the planar one can
accommodate black hole solutions of the vacuum Einstein equations. In what follows, we will show that there effectively exists a black hole solution with hyperscaling violation of the Einstein equations with a Nil geometry horizon. In addition, we will show that two other non-trivial geometries (the Solv geometry and the $\mbox{SL}_2(\mathds{R})$) are not suitable even for zero mass solutions of the vacuum Einstein equations.

%%%%%%%%%%%%%%%%%%%%%%%%%%%%%%%%%%%%%%%%%%%%%%%%%%%%%%%%%%%%%%%%%%%%%%%%%%%%%%%%%%%%%%%%%%%%%%%%%%%%%%%%
\section{Black hole of vacuum Einstein equations with hyperscaling violation and Nil geometry horizon}
%%%%%%%%%%%%%%%%%%%%%%%%%%%%%%%%%%%%%%%%%%%%%%%%%%%%%%%%%%%%%%%%%%%%%%%%%%%%%%%%%%%%%%%%%%%%%%%%%%%%%%%%
Let us consider the Einstein-Hilbert action in five dimensions
\begin{eqnarray}
S=\frac{1}{2\kappa}\int \sqrt{-g}\,d^5x\, R
\end{eqnarray}
where $\kappa$ is proportional to the Newton coupling constant and $R$ stands for the scalar curvature. The vacuum field equations are equivalent to
\begin{eqnarray}
R^{\mu}_{\,\,\nu}=0,
\label{VEeqs}
\end{eqnarray}
where $R_{\mu\nu}$ denotes the Ricci tensor. We look for a black hole solution with hyperscaling violation with a base manifold modeled by one the three non-trivial Thurston geometries (\ref{Thurstongeo}) within the following ansatz
\begin{eqnarray}
\frac{1}{r^{\frac{2\theta}{3}}}\left[-{r^{2z}}F(r){dt^2}+\frac{dr^2}{r^2 F(r)}+\sum_{I=1}^{3}\,r^{2q_I}\left({\omega}^I\right)^2\right].
\label{ansatz}
\end{eqnarray}
Here the single metric function $F(r)$ is assumed to have a zero corresponding to the location of the horizon, and in order to correctly reproduce the asymptotic behavior, we also impose that $\lim_{r\to\infty}F=1$. Note that since we are dealing without cosmological constant the scaling constants $a_I$ can be taken to unit without any loss of generality.

First, we consider the zero-mass case (that is for $F(r)=1$) with the choice of the exponents $q_I$ that ensure the metric to be homogeneous. Let us denote by $\tilde{R}^I_{\,\,J}$, the Ricci tensor relative of the three Thurston geometries (\ref{Thurstongeo}). A simple computation shows that in the homogeneous case, the components of the Ricci tensor for the zero mass metric (\ref{ansatz}), that is with $F(r)=1$ are generically given  by
\begin{subequations}
\begin{eqnarray}
\label{eqtt}
R^t_{\,\,t}=-r^{\frac{2\theta}{3}}\left(z-\frac{\theta}{3}\right)\xi,\\
\label{eqrr}
R^r_{\,\,r}=-r^{\frac{2\theta}{3}}\left[\sum_I q_I^2+z^2-\frac{\theta}{3}(\theta+\xi)\right],\\
\label{eqii}
R^I_{\,\,J}=r^{\frac{2\theta}{3}}\left[\tilde{R}^I_{\,\,J} \left(1+(q_{\hat{I}}-q_{\hat{J}})\xi\right)-\delta^I_{\,\,J}(q_{\hat{J}}-\frac{\theta}{3})\xi\right],
\end{eqnarray}
\end{subequations}
where for simplicity we have defined $\xi=\sum_I q_I+z-\theta$, and used the convention that indices with hat are not summed. The homogeneous conditions as well as
the components of $\tilde{R}^I_{\,\,J}$ are given in the following table for each of the three non-trivial Thurston geometries (\ref{Thurstongeo})
\begin{widetext}
\begin{center}
\begin{table}[h!]
\begin{center}
\begin{tabular}{|c|c|c|}
\hline
Horizon  & homog. cond. &  Non zero comp. of $\tilde{R}^I_{\,\,J}$\\
\hline \hline
Solv & $q_1=q_2, q_3=0$ & $ -2\delta^I_{\,\,x3}\delta^{x3}_{\,\,J}$\\
\hline \hline
Nil & $q_3=q_1+q_2$  & $\tilde{R}^i_{\,\,j}=-1/2\,\delta^i_{\,\,j}$,\quad $\tilde{R}^{x3}_{\,\,x3}=1/2$,\quad $\tilde{R}^{x3}_{\,\,x2}=-x1$ \\
\hline\hline
\mbox{SL}$_2(\mathds{R})$ & $q_I=0$ &  $\tilde{R}^i_{\,\,j}=-3/2\,\delta^i_{\,\,j}$,\quad $\tilde{R}^{x3}_{\,\,x3}=1/2$,\quad $\tilde{R}^{x3}_{\,\,x2}=2/x1$\\
\hline\hline
\end{tabular}
\caption{ {Homogeneous conditions and Ricci scalar components of the Thurston geometries} }
\end{center}
\end{table}
\end{center}
\end{widetext}
From the table I, it is clear that in all the cases $\theta=3z$ while the parameter $\xi=\sum_I q_I+z-\theta$ must be non-vanishing, $\xi\not=0$. For the Solv geometry ansatz, the equation (\ref{eqii}) for $I=J=i$ will imply that $q_i=\frac{\theta}{3}=z$, and because of the homogeneity condition $q_3=0$, this would imply that $\xi=0$ which is not a compatible condition as said before; hence the Solv geometry can not be solution of the vacuum Einstein equations. In the case of \mbox{SL}$_2(\mathds{R})$, we end with the same conclusion since the equation (\ref{eqii}) in the homogeneous case $q_I=0$
becomes $\tilde{R}^I_{\,\,J}+\frac{\theta (z-\theta)}{3}\delta^I_{\,\,J}=0$, and this equation is clearly incompatible with the form of the expressions of the components $\tilde{R}^I_{\,\,J}$. The case of the Nil geometry can be viewed as a mix of the two previous cases in the sense that the homogeneous conditions are not so restrictive as for the Solv geometry, and in the other hand, there is enough non-zero components  $\tilde{R}^I_{\,\,J}$ as it occurs for the \mbox{SL}$_2(\mathds{R})$ case. These are precisely these two
features that allow the Nil geometry metric with hyperscaling violation, that is the metric (\ref{ansatz}) with $F=1$, to be solution of the vacuum Einstein equations with dynamical exponent $z=3/2$ and hyperscaling violation exponent $\theta=9/2$. Moreover, this solution can be promoted to a static black hole of the vacuum Einstein equations (\ref{VEeqs}) with hyperscaling violation and with Nil geometry horizon whose line element is given by
\begin{widetext}
\begin{eqnarray}
ds^2=\frac{1}{r^3}\left[-r^3\Big(1-\frac{r_h}{r}\Big)dt^2+\frac{dr^2}{r^2\left(1-\frac{r_h}{r}\right)}+r^2(dx_1^2+dx_2^2)+r^4(dx_3-x_1\,dx_2)^2\right],
\label{bhmetric}
\end{eqnarray}
\end{widetext}
where the constant $r_h$ denotes the location of the horizon. One may note that this solution has exactly the same dynamical exponent $z=3/2$ than the Nil solution of Cadeau and Woolgar in presence of a cosmological constant (\ref{NilCW}). This analogy can be made more explicitly by rewriting the line element (\ref{bhmetric}) in the following suggestive form
\begin{widetext}
\begin{eqnarray}
ds^2=\frac{1}{r^{\frac{2\theta}{3}}}\left[-r^{2z}\Big(1-\frac{r_h}{r^{\sum_I q_I+z-\theta}}\Big)dt^2+\frac{dr^2}{r^2\left(1-\frac{r_h}{r^{\sum_I q_I+z-\theta}}\right)}+r^2(dx_1^2+dx_2^2)+(\frac{11}{2}-\theta)r^4(dx_3-x_1\,dx_2)^2\right].
\label{bhmetric2}
\end{eqnarray}
\end{widetext}
It is simple to see that for $z=3/2$, $\theta=9/2$, $q_i=1$ and $q_3=2$, this line element reduces to the solution (\ref{bhmetric}). But what it is interesting with this expression is that,  in the limit $\theta\to 0$, it exactly reproduces the Nil solution of Cadeau and Woolgar with $z=3/2$, see formulas (\ref{CW}, \ref{NilCW}, \ref{genCW}).

We now turn to the thermodynamics analysis. The partition function for a thermodynamical ensemble is identified
with the Euclidean path integral in the saddle point approximation
around the Euclidean continuation of the classical solution
\cite{Gibbons:1976ue}. The Euclidean and Lorentzian action are
related by $I_{E}=-iI$ where the periodic Euclidean time is  $\tau
=it$. The Euclidean continuation of the class of metrics considered here  is given by
\begin{eqnarray*}
ds^2=&&N(r)^2f(r)d\tau^2+\frac{dr^2}{f(r)}\nonumber\\
&&+\frac{1}{r^{2\theta/3}}\left[r^2(dx_1^2+dx_2^2)+r^4(dx_3-x_1\,dx_2)^2\right].
\end{eqnarray*}
In order to avoid conical singularity at the horizon in the
Euclidean metric, the Euclidean time is made periodic with period
$\beta$ related to the Hawking temperature $T$ by $T=\beta^{-1}$.
Here we are interested only in the static solution with a radial dependence, and hence it is enough to consider a
\textit{reduced} action principle given in this case
\begin{widetext}
\begin{eqnarray}
\label{redaction}
I_E=\frac{\beta\vert\Omega_3\vert}{2\kappa}\int_{r_h}^{\infty}dr\,N\left[(\theta-4) r^{3-\theta}f^{\prime}-\frac{2}{3}r^{2-\theta}(2\theta^2-13\theta+21)f-\frac{1}{2}r^{4-\frac{\theta}{3}}\right]+B,
\end{eqnarray}
\end{widetext}
where the radial coordinate $r$ belongs to the range $[r_h,\infty[$, the volume element of the three-dimensional horizon is denoted by $\vert\Omega_3\vert$ and
$B$ is a boundary term that is fixed by requiring that the Euclidean action has an extremum, that is $\delta I_{E}=0$. This in turn implies that
$$
\delta B=-\frac{\beta\vert\Omega_3\vert}{2\kappa}(\theta-4)\left[N\,r^{3-\theta}\left(\delta f\right)\right]_{r=r_h}^{r=\infty}.
$$
For the black hole Nil solution (\ref{bhmetric}), that is for $z=3/2$, $\theta=9/2$ and the metric functions given by
\begin{eqnarray}
f(r)=r^5\left(1-\frac{r_h}{r}\right),\qquad N(r)=r^{-\frac{5}{2}},
\label{redf}
\end{eqnarray}
the reduced action $I_E$ (\ref{redaction}) becomes
\begin{eqnarray}
I_E=\beta\left(\frac{1}{4\kappa}\vert\Omega_3\vert\, r_h\right)+\frac{2\pi}{\kappa \sqrt{r_h}}\vert\Omega_3\vert.
\end{eqnarray}
Since, in the grand canonical ensemble, the Euclidean action is related to the mass ${\cal M}$ and entropy ${\cal S}$ by
\begin{eqnarray}
I_E=\beta {\cal M}-{\cal S},
\label{gce}
\end{eqnarray}
one can easily identify the mass and entropy to be
\begin{eqnarray}
{\cal M}=\frac{1}{4\kappa}\vert\Omega_3\vert\, r_h,\quad {\cal S}=-\frac{2\pi}{\kappa \sqrt{r_h}}\vert\Omega_3\vert, \quad T=\frac{1}{4\pi}r_h^{\frac{3}{2}},
\label{thermo}
\end{eqnarray}
where $T$ is the  Hawking temperature. Curiously enough, in spite of the fact that the entropy is negative, the first law of thermodynamics holds,
i. e. $d{\cal M}=Td{\cal S}$, and the mass is positive.  Another curiosity is the fact that the reduced action (\ref{redaction}) possesses a scaling symmetry given by
\begin{eqnarray*}
\bar{r}=\sigma r,\quad \bar{N}(\bar{r})=\sigma^{-5+\frac{\theta}{3}}N(r),\quad \bar{f}(\bar{r})=\sigma^{2(1+\frac{\theta}{3})}f(r),
\end{eqnarray*}
from which one can derive a Noether conserved charge given by
\begin{eqnarray*}
{\cal C}(r)=\frac{\beta\vert\Omega_3\vert}{2\kappa}N(\theta-4) r^{3-\theta}\left(rf^{\prime}-2(1+\frac{\theta}{3})f\right).
\end{eqnarray*}
This latter yields for the solution (\ref{redf}) with $z=3/2$ and $\theta=9/2$ to ${\cal C}(r)=\beta {\cal M}$.

One natural question to ask is wether this kind of solution can be extended in dimension $D>5$. The answer is positive, and the pattern to obtain black hole solutions with hyperscaling violation and Nil horizon is quite similar to the one described previously in five dimensions. For example in six dimensions, the Nil $4-$geometry has the following left-invariant one-forms
\begin{eqnarray*}
\omega^1=dx_1,\, \omega^2=dx_2-x_3dx_1,\, \omega^3=dx_3-x_4dx_1,\,\omega^4=dx_4,
\end{eqnarray*}
and the the static black hole with hyperscaling violation and Nil $4-$ horizon topology is given by
\begin{eqnarray}
\frac{1}{r^{\frac{\theta}{2}}}\left[-{r^{2z}}F(r){dt^2}+\frac{dr^2}{r^2 F(r)}+\sum_{I=1}^{4}\,r^{2q_I}\left({\omega}^I\right)^2\right].
\label{Nil 6}
\end{eqnarray}
with the following set of parameters
\begin{eqnarray}
&&z=\frac{3\sqrt{2}}{2},\,q_1=\frac{\sqrt{2}}{2},\,q_2=2\sqrt{2},\,q_3=\frac{3\sqrt{2}}{2},\, q_4=\sqrt{2}\nonumber\\
&&\theta=6\sqrt{2},\qquad F(r)=1-\frac{M}{r^{\frac{\sqrt{2}}{2}}}.
\end{eqnarray}
As in the five-dimensional case, the metric function presents the same structure, namely
$$
F(r)=1-\frac{M}{r^{\sum_Iq_I+z-\theta}}.
$$

%%%%%%%%%%%%%%%%%%%%%%%
\section{Conclusions}
%%%%%%%%%%%%%%%%%%%%%%%
Here, we have considered the vacuum Einstein equations in five dimensions for which we have obtained a new static black hole solution with hyperscaling violation and where the $3-$dimensional horizon is modeled by the Nil geometry. In this case, the dynamical exponent $z$ is fixed as $z=3/2$ as well as the hyperscaling exponent $\theta=3z$. Along the same line, we have shown that the two others non-trivial Thurston geometries, namely the Solv geometry and SL$_2(\mathds{R})$, can not accommodate a similar construction. Using the Hamiltonian formalism, we have computed the mass and entropy of the Nil black hole solution. Curiously enough, in spite of the fact that the entropy is negative, the mass is positive and the first law of thermodynamics still holds.

An interesting work will be to explore the extension in higher dimension. This can be achieved in different ways. One can consider the extension of the Nil geometries in dimension greater than three as we did in the end of the last section. Another option will be to consider as possible horizons, the generalized Heisenberg spaces \cite{Hervik:2003vx}. There is also the option of considering arbitrary product of Nil, Solv, SL$_2(\mathds{R})$ or generalized Heisenberg geometries as it has been done for example in the case of Born-Infeld gravity with cosmological constant in \cite{Anabalon:2011bw}.

The charged and/or the spinning version of the solution found here are also two interesting but non-trivial tasks.

%%%%%%%%%%%%%%%%%%%%%%%%%%%%%%%%%%%%%%%%%%%%%%%%%%%%%%%%%%%%%%%%%%%%%%%%%%%%%%%%%%%%%%%%%%%%%%%%
\begin{acknowledgments}
We thank Eloy Ay\'on-Beato, Moises Bravo and Julio Oliva for useful discussions.  This work is partially supported by grant
1130423 from FONDECYT and from
CONICYT, Departamento de Relaciones Internacionales ``Programa
Regional MATHAMSUD 13 MATH-05''. This project was partially funded by Proyectos
CONICYT- Research Council UK - RCUK -DPI20140053.
\end{acknowledgments}
%%%%%%%%%%%%%%%%%%%%%%%%%%%%%%%%%%%%%%%%%%%%%%%%%%%%%%%%%%%%%%%%%%%%%%%%%%%%%%%%%%%%%%%%%%%%%%%%%%

%%%%%%%%%%%%%%%%%%%%%%%%%%%

%%%%%%%%%%%%%%%%%%%%%%%%%%%

\end{document}